\documentclass[epj]{svjour}
\usepackage{amssymb}
\usepackage{graphics}
\begin{document}
\title{Interaction-enhanced double resonance in cold gases}

\author{A. I. Safonov\inst{1}, I. I. Safonova\inst{1} and I. S. Yasnikov\inst{2}}
\institute{National Research Centre Kurchatov Institute, Moscow 123182, Russia \and Togliatti
State University, Togliatti 445667, Russia}
\date{Received: January 14, 2011 / Revised version: April 5, 2011}
%
\abstract{A new type of double-resonance spectroscopy of a quantum gas based on
interaction-induced frequency modulation of a probe transition has been considered. Interstate
interaction of multilevel atoms causes a coherence-dependent collisional shift of the transition
between the atomic states $|1\rangle$ and $|2\rangle$ due to a nonzero population of the state
$|3\rangle$. Thus, the frequency of the probe transition $|1\rangle-|2\rangle$ experiences
oscillations associated with the Rabi oscillations between the states $|1\rangle$ and $|3\rangle$
under continuous excitation of the drive resonance $|1\rangle-|3\rangle$. Such a dynamic frequency
shift leads to a change in the electromagnetic absorption at the probe frequency and,
consequently, greatly enhances the sensitivity of double-resonance spectroscopy as compared to
traditional ``hole burning'', which is solely due to a decrease in the population of the initial
state $|1\rangle$. In particular, it has been shown that the resonance linewidth is determined by
the magnitude of the contact shift and the amplitude of the drive field and does not depend on the
static field gradient. The calculated line shape and width agree with the low-temperature
electron-nuclear double-resonance spectra of two-dimensional atomic hydrogen.}

\maketitle
\section{Introduction}
\label{sec:intro} Interaction of atoms and, more specifically, short-range or contact interaction,
leads to a shift of internal atomic transitions commonly referred to as a contact shift. In the
case of multi-level atoms, the contact shift also depends on interstate coherence and populations
of the states not involved in the particular resonance. More specifically, according to
experiments on $^6$Li~\cite{Gupta} and $^{40}$K~\cite{Regal_Jin} and recent
theory~\cite{Baym,Saf_JLTP}, an incoherent population of such a third state causes a nonzero shift
for fermions. On the other hand, fermions in identical internal states do not interact via $s$
waves; therefore, a fully coherent ultra-cold Fermi gas exhibits a contact shift only in the case
of spatial inhomogeneity~\cite{Campbell,Gibble}. On the contrary, the respective contribution to
the contact shift of bosons is the highest if the internal states are coherently populated and
decreases by a factor of 2 as the sample decoheres. Thus, in a Bose gas, excitation of a drive
transition to the state $|3\rangle$ induces frequency modulation of the probe transition
$|1\rangle-|2\rangle$ associated with Rabi oscillations between the states $|1\rangle$ and
$|3\rangle$~\cite{Saf_JLTP}. The latter leads to interaction-enhanced double resonance when two
transitions are excited simultaneously. In this work, we consider this effect in the
relaxation-free approximation, analyze a possible double-resonance lineshape and compare it with
experiments in spin-polarized atomic hydrogen~\cite{Ahokas,Ahokas_JLTP}.

Interaction of multilevel systems with several resonance fields is commonly considered with the
use of the evolution equations for the components of spin density matrix~\cite{CPT}. Inclusion of
the population-dependent frequency shift makes these equations essentially nonlinear and their
analytical solution impossible. Our approach is less comprehensive but provides quantitative and
physically clear description in the case of interest.

\section{Interaction-enhanced double resonance}
\label{sec:inedor}
\begin{figure}
\resizebox{\columnwidth}{!}{\includegraphics{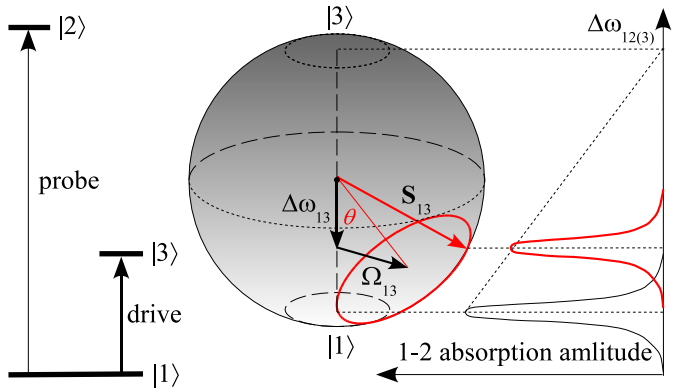}} \caption{(Left) Three-level
scheme and (right) effect of Rabi oscillations between the states $|1\rangle$ and $|3\rangle$ on
the $|1\rangle-|2\rangle$ transition frequency and intensity. Here, for better visibility, the
Rabi period $2\pi/\Omega_{13}$ is thought to be much longer than the time needed to detect the
$|1\rangle-|2\rangle$ resonance line. The opposite case is considered in Secs. \ref{sec:spectrum}
and \ref{sec:H} below.} \label{fig:Bloch}
\end{figure}

Physics of this novel phenomenon becomes transparent already in the case of just
three levels (Fig.~\ref{fig:Bloch}). In general, excitation of the probe transition
$|1\rangle-|2\rangle$ affects the frequency of the drive $|1\rangle-|3\rangle$ resonance and vise
versa. This may lead to specific effects and produce peculiar double-resonance spectra, which will
be considered elsewhere.

Here, for simplicity and clarity, we restrict ourselves to the case (see below for the exact
condition) when the contact shift vanishes in the absence of the third state. Another advantage of
this case is that it allows direct comparison with experiments on electron-nuclear double
resonance (ENDOR) in atomic hydrogen~\cite{Ahokas,Ahokas_JLTP}. To simplify our consideration even
further, we assume that the population of the state $|2\rangle$, which is the final state of the
probe transition, is kept negligibly small and therefore the frequency of the drive transition
remains constant. On the other hand, the frequency shift of the probe transition
$|1\rangle-|2\rangle$ in the presence of the state $|3\rangle$ is~\cite{Saf_JLTP}
\begin{eqnarray}
\hbar\Delta\omega^{\rm Bose}_{12(3)}&=&2n_1(\lambda^+_{12}-\lambda_{11})+2n_2(\lambda_{22}-\lambda^+_{12})\nonumber\\
&+&2n_3|C^+_{13}|^2(\lambda^+_{23}-\lambda^+_{13}),\label{eq:Dom123_b}\\
\hbar\Delta\omega^{\rm
Fermi}_{12(3)}&=&2n_3|C^-_{13}|^2(\lambda^-_{23}-\lambda^-_{13}).\label{eq:Dom123_f}
\end{eqnarray}
Here, $n_i$ is the population of the state $|i\rangle$, $\lambda^\pm_{ij}\equiv\langle
ij|\lambda|ij\rangle_\pm$ is the matrix element (excluding the spatial factor) of the interaction
strength $\lambda=4\pi\hbar^2a/m$ commonly used in a cold-collision regime, $m$ is the atomic
mass, and $a$ is the respective scattering length. $C_{ij}^\pm$ are the normalized
($|C_{ij}^+|^2+|C_{ij}^-|^2=1$) amplitudes of the (+) symmetric and ($-$) antisymmetric component
of properly symmetrized diatomic wavefunction $ |{\bf k}i,{\bf q}j\rangle =
C_{ij}^+\psi_\pm|ij\rangle_+ + C_{ij}^-\psi_\mp|ij\rangle_-$, where the upper (lower) subscript
stands for bosons (fermions) and the internal (pseudo-spin) and spatial parts are, respectively,
$|ij\rangle_\pm=\frac{1}{\sqrt{2}}(|ij\rangle\pm|ji\rangle)$ and
$\psi_\pm=\frac{1}{\sqrt{2}}(\psi_{\bf k}({\bf r}_1)\psi_{\bf q}({\bf r}_2)\pm\psi_{\bf q}({\bf
r}_1)\psi_{\bf k}({\bf r}_2))$.

Equation~(\ref{eq:Dom123_f}) should be compared with Eq.(1) of Gupta {\it et al.}~\cite{Gupta} and
Eq. (1) of Regal and Jin~\cite{Regal_Jin}, which hold in a fully incoherent case of
$|C^-_{13}|^2=\frac{1}{2}$, as well as with the general theoretical result (Eqs. (8)-(9) of Baym
{\it et al.}~\cite{Baym}). Equation (\ref{eq:Dom123_b}) generalizes the two-level formula (4) of
Gibble~\cite{Gibble}. Clearly, the shift in the two-level gas ($n_3 = 0$) is zero for fermions and
for bosons with $\lambda_{11}=\lambda^+_{12}=\lambda_{22}$. The latter is exactly the case
mentioned above and agrees with our previous result~\cite{Safonov_08} and with the observations in
atomic hydrogen~\cite{Ahokas,Ahokas_3D}. Moreover, in agreement with other
authors~\cite{Gupta,Gibble,Harber,Zwierlein}, the shift is independent of the sample coherence,
because the fraction $|C_{12}^-|$ of pseudo-singlets does not enter Eqs. (\ref{eq:Dom123_b}) and
(\ref{eq:Dom123_f}). Thus, we come to the well-known conclusion that the interstate coherence in a
spatially homogeneous two-level gas cannot be probed by the contact shift~\cite{Gibble}. Note,
that spatial inhomogeneity renders the atoms distinguishable and causes specific
coherence-dependent shifts of opposite signs for bosons and fermions~\cite{Campbell,Gibble}.

The situation in a three-level gas ($n_3\neq0)$ is fundamentally different. We see that the
interstate coherence of a spatially homogeneous gas can be probed by the contact shift in the
presence the third state, like it has been done in fermionic $^6$Li~\cite{Gupta}. It is also
remarkable that a homogeneous Fermi gas does exhibit a nonzero clock shift if the two states
involved in the transition are not fully coherent with the third one ($C^-_{13}\neq0$), the latter
is populated ($n_3\neq0$) and the respective scattering lengths are different ($a^-_{13}\neq
a^-_{23}$). In this context, an incoherently populated third state acts as a buffer gas of foreign
atoms. One can also see that, like in the case of just two levels, the coherence between the
one-atom states $|1\rangle$ and $|2\rangle$ coupled by the resonance transition does not affect
the clock shift $\Delta\omega_{12(3)}$ of this resonance.

In the absence of relaxation, the populations of the states $|1\rangle$ and $|3\rangle$ evolve
(see Fig.~\ref{fig:Bloch}) under continuous excitation of the transition $|1\rangle-|3\rangle$ in
the initially $|1\rangle$-state sample as
$n_3=n\sin^2\theta_{13}\sin^2\frac{\tilde{\Omega}_{13}t}{2}$, $n_1=n-n_3$, where $n$ is the total
density of the gas, $\sin\theta_{13}=\Omega_{13}/\tilde{\Omega}_{13}$,
$\tilde{\Omega}_{13}=\sqrt{\Delta\omega_{13}^2+\Omega_{13}^2}$, $\Omega_{13}$ is the respective
Rabi frequency, and $\Delta\omega_{13}=\omega_{\rm d}-\omega_{13}$ is the detuning between the
resonance frequency $\omega_{13}$ and the frequency $\omega_{\rm d}$ of the drive field. Note,
that the gas remains fully coherent, since pseudo-spins $S_{13}$ of all atoms experience
\emph{coherent} precession on the $|1\rangle-|3\rangle$ Bloch sphere around the tilting angle
$\theta_{13}$ and stay parallel to each other (Fig. \ref{fig:Bloch}). Thus, the resonance
frequency $\omega_{13}$ remains unchanged while the $|1\rangle-|2\rangle$ transition is
dynamically shifted to the frequency~\cite{Saf_JLTP}
\begin{equation}\label{eq:bc_dyn}
\hbar\omega_{12}=\hbar\omega_{12}^{(0)}+2n\Delta\lambda\sin^2\theta_{13}\sin^2\frac{\tilde{\Omega}_{13}t}{2},
\end{equation}
where $\Delta\lambda=\lambda^+_{23}-\lambda^+_{13}$. The amplitude of this frequency modulation
can easily be comparable with or even much greater than the linewidth of the probe transition.
Thus, excitation of the transition $|1\rangle-|3\rangle$ periodically drives the probe transition
out of resonance. For clarity, Fig.~\ref{fig:Bloch} corresponds to slow driving in a sense that
the Rabi period $2\pi/\Omega_{13}$ of the drive transition is thought to be much longer than the
time needed to detect the $|1\rangle-|2\rangle$ resonance line. The opposite case of fast driving
is considered in Secs. \ref{sec:spectrum} and \ref{sec:H} below. At slow driving, the entire
$|1\rangle-|2\rangle$ spectrum moves periodically fore and back along the frequency axis and
simultaneously changes in amplitude. The waveforms shown in Fig.~\ref{fig:Bloch} are simply the
snapshots of the spectrum at different phases of the Rabi cycle. Obviously, electromagnetic
absorption at the fixed frequency $\omega_{\rm p}$ of the probe field also changes periodically
with a period of the $|1\rangle-|3\rangle$ Rabi oscillations. It should be emphasized that these
changes are associated with changes in \textit{both} the population of the initial state
\textit{and} the transition frequency, in contrast to conventional double resonance, which is
solely due to a change in the population of the initial state caused by the drive transitions.
Clearly, such interaction-induced frequency modulation can greatly enhance the effect, which
therefore can be called INteraction-Enhanced DOuble Resonance (INEDOR).

\section{INEDOR spectrum}
\label{sec:spectrum} Let us consider in more detail a possible line shape of the INEDOR spectrum.
In contrast to the case of slow driving illustrated in Fig. 1, we assume that simultaneous
frequency and amplitude modulation of the $|1\rangle-|2\rangle$ absorption line is relatively fast
(which is the case in a sufficiently high drive field) and therefore integrated by the detection
system. This implies that the $|1\rangle-|3\rangle$ Rabi frequency is much higher than the rate of
field or frequency sweep through the $|1\rangle-|2\rangle$ resonance or the inverse time constant
$\tau^{-1}$ of the detection system, $\Omega_{13}\tau\gg1$. In practice, as discussed in
Sec.~\ref{sec:H} below, a less severe condition $\tilde{\Omega}_{13}\tau\gg1$ is sufficient. Thus,
we deal with time-average absorption at the probe frequency in a sense that the absorption signal
is integrated over many modulation (Rabi) cycles.

\begin{figure}
\resizebox{\columnwidth}{!}{\includegraphics{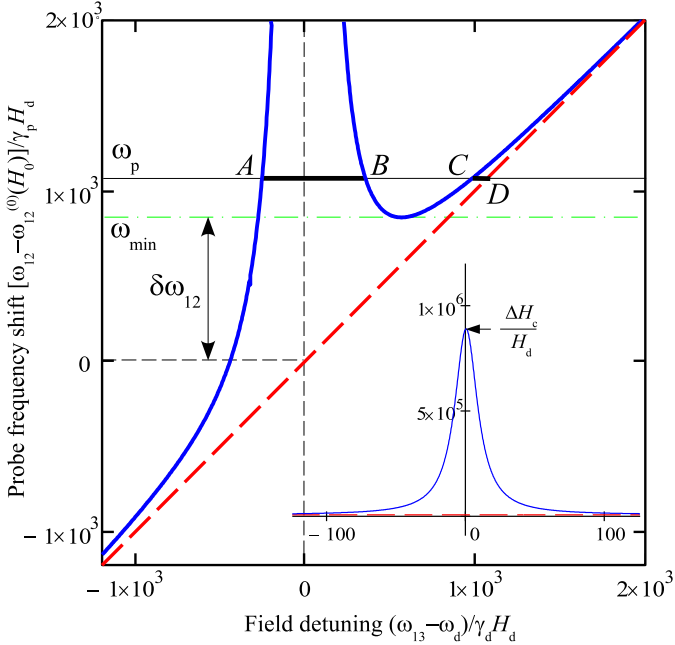}} \caption{Field dependence of the
$|1\rangle-|2\rangle$ frequency shift (in units of $\gamma_{\rm p}H_{\rm d}$) under the CW
excitation of the $|1\rangle-|3\rangle$ resonance. Horizontal axis is the field detuning
$(\omega_{13}-\omega_{\rm d})/\gamma_{\rm d}$ from the $|1\rangle-|3\rangle$ resonance in units of
the excitation field $H_{\rm d}$. Solid and dashed line are, respectively, the upper bound of the
sum of the Zeeman and mean-field contributions and the Zeeman contribution alone. Horizontal lines
indicate (dash-dotted line) the value of $\omega_{12}$ at minimum and (solid line) the probe
frequency. Vertical dashed line corresponds to the resonance field for the $|1\rangle-|3\rangle$
transition. The parameters correspond to 2D atomic hydrogen with a density of $3\cdot10^{12}$
cm$^{-2}$ in a high polarizing field of 45 kG, except for the sign of the contact shift: $H_{\rm
d}=1$~mG, $\Delta H_{\rm c}=89$~G, $\gamma_{\rm p}=\gamma_e$, $\gamma_{\rm d}=\gamma_p$. Inset is
the overview of the same dependence.} \label{fig:Frequency}
\end{figure}

Generally, the energies of all three levels and, consequently, both transition frequencies depend
on the external static field. The nature of this field is insignificant but for definiteness we
shall consider the magnetic field $H$. Since our analysis is conducted in terms of pseudo-spins,
the respective contribution to the frequencies is the Zeeman splitting $\gamma_{\rm d(p)}H$, where
$\gamma_{\rm d(p)}$ is the effective gyromagnetic ratio of the drive (probe) transition. Let the
field $H_0$ correspond to the exact $|1\rangle-|3\rangle$ resonance, $\omega_{13}(H_0)=\omega_{\rm
d}$. Then the deviation $h=H-H_0$ of the field from this value results in the drive frequency
detuning $\Delta\omega_{13}(h)=\gamma_{\rm d}h$. Owing to this Zeeman contribution to the
effective precession frequency $\tilde{\Omega}_{13}$, the amplitude of the oscillating component
of $\omega_{12}$ in Eq.~(\ref{eq:bc_dyn}), i.e., the amplitude of the frequency modulation of the
probe transition turns out be a Lorentzian function of the field
\begin{equation}\label{eq:mod_ampl}
\Delta\omega_{12}(h)=\left(\frac{2n\Delta\lambda}{\hbar}\right)\frac{H_{\rm d}^2}{H_{\rm
d}^2+h^2},
\end{equation}
where $H_{\rm d}$ is the amplitude of the drive field $H_{\rm d}(t)=H_{\rm d}\exp(i\omega_{\rm
d}t)$, $\Omega_{13}=\gamma_{\rm d}H_{\rm d}$. On the other hand, the static (zero-density)
component of $\omega_{12}$ also includes the Zeeman term
\begin{equation}\label{eq:omega12_stat}
\omega_{12}^{(0)}(H_0+h)=\omega_{12}^{(0)}(H_0)+\gamma_{\rm p}h.
\end{equation}
The sum of these two contributions is an upper bound of the probe frequency,
\begin{equation}\label{eq:12_tot}
\omega_{12}=\omega_{12}^{(0)}(H_0)+\gamma_{\rm
p}h+\left(\frac{2n\Delta\lambda}{\hbar}\right)\frac{H_{\rm d}^2}{H_{\rm d}^2+h^2}.
\end{equation}
As a result, $\omega_{12}(t)$ oscillates between the Zeeman-only lower bound
(Eq.~(\ref{eq:omega12_stat}), dashed line in Fig.~(\ref{fig:Frequency})) and the Zeeman plus
mean-field upper bound (Eq.~(\ref{eq:12_tot}), thick solid line in Fig.~(\ref{fig:Frequency})) at
the field-dependent frequency $\tilde{\Omega}_{13}(h)=\gamma_{\rm d}\sqrt{H_{\rm d}^2+h^2}$. As is
seen from Eq.~(\ref{eq:12_tot}), the upper bound of the probe frequency is generally a
nonmonotonic function of the static field.

As mentioned above, the Rabi frequency $\Omega_{13}=\tilde{\Omega}_{13}(0)$ is assumed to be
sufficiently high, $\Omega_{13}\tau\gg1$, in which case the absorption amplitude is integrated by
the detection system over many Rabi cycles. The time-average absorption amplitude $A(h,
\omega_{\rm p})$ within the bounds (\ref{eq:omega12_stat}) and (\ref{eq:12_tot}) is proportional
to the average population of the initial state multiplied by the probability density $\rho(h,
\omega_{\rm p})=[\tilde{\Omega}_{13}(h)/2\pi][d\omega_{12}(h, \omega_{\rm p})/dt]^{-1}$ to find
the system at given $h$ and $\omega_{\rm p}$. It is readily shown that
\begin{equation}\label{eq:Amplitude}
A(h, \omega_{\rm p})\propto\frac{1-x}{2\pi\sqrt{x(\sin^2\theta_{13}(h)-x)}},
\end{equation}
where
\begin{eqnarray*}
x=\frac{\hbar[\omega_{\rm p}-\omega_{12}^{(0)}(H_0+h)]}{2n\Delta\lambda}
\end{eqnarray*}
and $1-x$ are the relative populations of the states $|3\rangle$ and $|1\rangle$, respectively.
Thus, the probability density peaks at the lower ($x=0$) and upper ($x=\sin^2\theta_{13}(h)$)
bound of the probe frequency. Note, that in the latter case absorption is further reduced due to a
smaller population of the initial state and vanishes at $x=1$ (which implies $h=0$), i.e., at the
maximum of the Lorentzian function (\ref{eq:mod_ampl}).

The spectrum that appears when both the drive and probe frequency are fixed and the field is
slowly swept through the resonance consists of up to four peaks corresponding to the field values,
at which the probe frequency shown by the solid horizontal line in Fig.~\ref{fig:Frequency}
crosses the upper (Zeeman plus mean-field) and lower (Zeeman only) bound of $\omega_{12}$ (points
$A$, $B$, $C$, and $D$). The actual number of lines depends on the amplitude of the contact shift,
the ratio $\gamma_{\rm d}/\gamma_{\rm p}$ of the Zeeman quanta of the excitation fields and the
relation between the drive and probe frequency. The peaks lying within the drive resonance (i.e.,
within the segment $AB$ in Fig.~\ref{fig:Frequency}) are much weaker because the atoms spend only
a small fraction of time in the resonance conditions. This implies that the $|1\rangle-|2\rangle$
absorption amplitude is strongly decreased, which produces a ``hole'' in the resonance curve.

So far we considered a spatially homogeneous system. There could be three types of inhomogeneity
associated with the inhomogeneous (i) static field, (ii) excitation fields and (iii) gas density.
In terms of the upper bound of the probe frequency shift in Fig.~\ref{fig:Frequency}, these
correspond to a spread in (i) the overall vertical position of the curve (equivalent to the
inhomogeneity of the probe frequency), (ii) the width of the drive resonance and (iii) the
amplitude of the drive resonance plus the overall vertical shift.

\begin{figure}
\resizebox{\columnwidth}{!}{\includegraphics{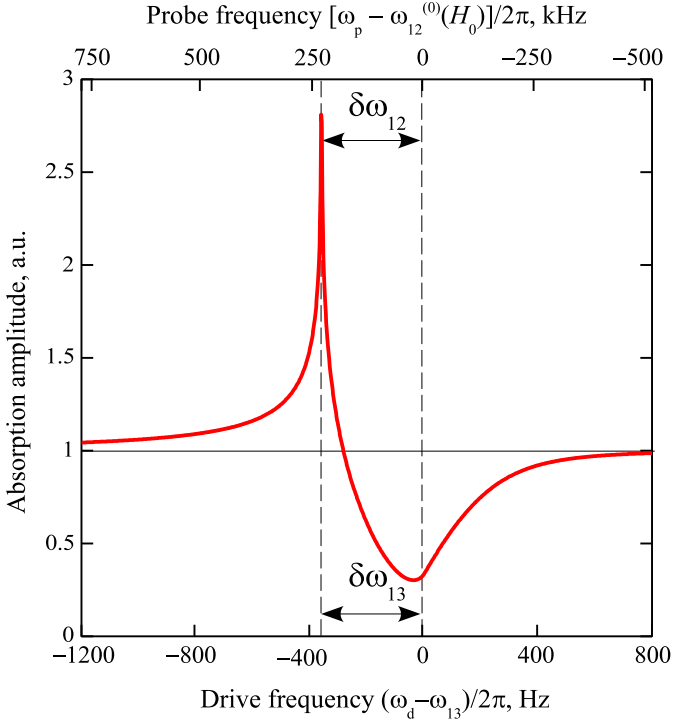}} \caption{$|1\rangle-|2\rangle$
absorption amplitude as a function of drive frequency (expressed as the detuning, in Hertz, from
the $|1\rangle-|3\rangle$ resonance) for the delta-functional spectrum of the probe field. Sharp
peak corresponds to the frequency minimum in Fig.~\ref{fig:Frequency}. Parameters are the same as
in Fig.~\ref{fig:Frequency}.} \label{fig:Spectrum}
\end{figure}

To illustrate the effect of inhomogeneity we consider in more detail a linear gradient $\nabla H$
of the static field in an otherwise spatially homogeneous infinite sample. In this case, all
possible field values are simultaneously present in different parts of the sample. Thus, the
absorption amplitude at a given probe frequency $\omega_{\rm p}$ is proportional to the integral
$I(\omega_{\rm p})=\int A(h, \omega_{\rm p})\left(\frac{\partial N}{\partial h}\right)dh$ of
Eq.~(\ref{eq:Amplitude}) along the $\omega_{\rm p}={\rm const}$ line within the bounds of the
probe frequency (i.e., over the horizontal segments $AB$ and $CD$ in Fig.~\ref{fig:Frequency}).
Owing to a constant field gradient, integration over the field is equivalent to integration over
space provided that the number of particles per unit volume $n$ is replaced by the number of
particles per unit field $\frac{\partial N}{\partial h}\propto n|\nabla H|^{-1}$. The result of
numerical integration with the parameters of experiments with 2D atomic
hydrogen~\cite{Ahokas,Ahokas_JLTP,Vas_priv} is shown in Fig.~\ref{fig:Spectrum} as a function of
the drive frequency $\omega_{\rm d}$ (lower horizontal axis) at constant $\omega_{\rm p}$.
Alternatively, the INEDOR spectrum can be detected by sweeping $\omega_{\rm p}$ (upper horizontal
axis) at constant $\omega_{\rm d}$. In the latter case, the spectrum is inverted on the frequency
scale because an increase in $\omega_{\rm d}$ corresponds to an increase in the resonance field
$\omega_{\rm d}/\gamma_{\rm d}$ and, consequently, to a positive displacement of the mean-field
Lorentzian peak on the $\omega_{12}(h)$ curve (\ref{eq:12_tot}) in Fig.~\ref{fig:Frequency}. The
latter is equivalent to a negative change in the probe frequency $\omega_{\rm p}$ at constant
$\omega_{\rm d}$. Which way of observing the INEDOR spectrum is preferred depends on the details
of a particular experiment.

The absorption amplitude as a function of $\omega_{\rm d}$ (Fig.~\ref{fig:Spectrum}) is nearly
constant well below and above the drive resonance, decreases substantially within the resonance
and has a sharp maximum at $\omega_{\rm p}=\omega_{\rm min}$, i.e., when the minimum value of
Eq.~(\ref{eq:12_tot}) coincides with the probe frequency. This explains the dispersion-looking
ENDOR spectra of 2D atomic hydrogen~\cite{Ahokas,Ahokas_JLTP} (see Sec.\ref{sec:H} below).
Physically, the hole in the absorption amplitude is because the atoms of the otherwise resonant
region of the sample are dynamically driven out of the probe resonance and spend only a small
fraction of time in the resonance conditions. As already mentioned, the probability of finding
them in resonance peaks at the lower (\ref{eq:omega12_stat}) and upper (\ref{eq:12_tot}) bound of
the probe frequency and, according to Eq.~(\ref{eq:Amplitude}), is the lower the higher is the
difference (\ref{eq:mod_ampl}) between the two bounds of $\omega_{12}$. In this respect, there is
no surprise that the positions of the $I$ minimum and the contact shift (\ref{eq:mod_ampl})
maximum nearly coincide, $\omega_{\rm p}\approx\omega^{(0)}_{12}(H_0)$. On the other hand, the
absorption maximum is due to the fact that the drive resonance introduces zero gradient of the
$|1\rangle-|2\rangle$ transition frequency in a certain region of the sample. As a result, much
more atoms become resonant. A decrease in the population of the initial state $|1\rangle$ at the
point of zero gradient turns out to be of minor importance for the chosen values of the
parameters. However, according to Eq.~(\ref{eq:12_tot}), it can be significant at
$|2n\Delta\lambda|\sim\hbar\gamma_{\rm p}H_{\rm d}$, which occurs at low gas density $n$ and
strong drive field $H_{\rm d}$.

The dispersion-looking INEDOR spectrum of a gas in a spatially inhomogeneous static field can be
also qualitatively explained in a different way. In the case of a positive contact shift, the
parts of the inhomogeneously broadened probe resonance line are shifted towards higher fields by
the amount proportional to the local amplitude of drive resonance curve. The shift increases with
an increase in field at the low-field side of the drive resonance and decreases at its high-field
side. Thus, the low-field side of the spectrum is ``rarefied'' whereas the high-field side is
``compressed'' in a sense that the number of atoms per unit field, which are involved in the probe
transitions at lower (higher) fields, becomes smaller (larger). Accordingly, the probe absorption
intensity is decreased at the low-field side of the drive resonance and increased at its
high-field side. When $\omega_{\rm d}$ is swept from below at constant $\omega_{\rm p}$, the
high-field side of drive resonance comes first. That is, the probe absorption intensity is
increased at lower and decreased at higher $\omega_{\rm d}$, as illustrated by
Fig.~\ref{fig:Spectrum}.

As discussed above, the width of the double-resonance curve in the probe-frequency units can be
estimated as the difference $\omega_{\rm min}-\omega^{(0)}_{12}(H_0)$ (see
Fig.~\ref{fig:Frequency}). For $|2n\Delta\lambda|\gg\hbar\gamma_{\rm p}H_{\rm d}$ the minimum of
the right-hand side of Eq.~(\ref{eq:12_tot}) occurs at
\begin{equation}\label{eq:minimum}
\frac{(H^2_{\rm d}+h^2)^2}{H^2_{\rm d}h}=\frac{4n\Delta\lambda}{\hbar\gamma_{\rm p}}.
\end{equation}
Typically, $h\gg H_{\rm d}$ and Eq.~(\ref{eq:minimum}) yields
\begin{equation}\label{eq:width_h}
h^3\simeq\frac{4n\Delta\lambda H^2_{\rm d}}{\hbar\gamma_{\rm p}}.
\end{equation}
This corresponds to
\begin{equation}\label{eq:width_w12}
\delta\omega_{12}\equiv\omega_{12}(h)-\omega^{(0)}_{12}(H_0)\simeq\frac{3}{2}\gamma_{\rm p}h
\end{equation}
and we finally arrive at
\begin{equation}\label{eq:width_w13}
\delta\omega_{13}=\frac{\gamma_{\rm d}}{\gamma_{\rm
p}}\delta\omega_{12}\simeq\frac{3}{2}\gamma_{\rm d}(2\Delta H_{\rm c}H^2_{\rm d})^{1/3},
\end{equation}
where $\Delta H_{\rm c}=2n\Delta\lambda(\hbar\gamma_{\rm p})^{-1}$ is the maximum contact shift of
the $|1\rangle-|2\rangle$ resonance in field units. Note, that the resonance linewidth depends on
the gas density and drive field as $n^{1/3}H_{\rm d}^{2/3}$. Remarkably, the width of the INEDOR
spectrum is independent of the static field gradient. On the other hand, the spectrum intensity is
inversely proportional to $|\nabla H|$. A one-dimensional inhomogeneity of the drive field in the
direction of $|\nabla H|$ also does not affect the linewidth because in this case $H_{\rm d}$ is a
single-valued function of $h$. The effect of such an inhomogeneity is that the mean-field peak
(\ref{eq:mod_ampl}) becomes asymmetric and the INEDOR linewidth is determined by the local $H_{\rm
d}$ value at the minimum of $\omega_{12}$ in Eq.~(\ref{eq:12_tot}). It is also worth mentioning
that the linearity of the Zeeman contributions to $\omega_{12}$ and $\omega_{13}$ is unnecessary
because these terms can be always linearized in the vicinity of the resonance field. In this case,
the gyromagnetic ratios should be replaced by the respective field derivatives, $\gamma_{\rm
d}\rightarrow
\partial\omega_{13}/\partial H$ and $\gamma_{\rm p}\rightarrow
\partial\omega_{12}/\partial H$.

\section{ENDOR in atomic hydrogen}
\label{sec:H}The spin states of a single hydrogen atom in a high magnetic field are commonly
referred to as $|a\rangle\approx|-+\rangle$, $|b\rangle=|--\rangle$, $|c\rangle\approx|+-\rangle$,
and $|d\rangle=|++\rangle$, in the order of increasing energy ($\pm$ signs stand for the
projections of the electron and nuclear spin on the field direction; we also neglect small
impurities of the opposite spin projections in the states $|a\rangle$ and $|c\rangle$). Ahokas
{\it et al.}~\cite{Ahokas,Ahokas_JLTP} observed a narrow feature in the ESR $|b\rangle-|c\rangle$
line while sweeping through the $|a\rangle-|b\rangle$ resonance. According to
Eq.~(\ref{eq:Dom123_b}), the $|b\rangle-|c\rangle$ resonance is dynamically shifted to the
frequency~\cite{Saf_JLTP}
\begin{equation}\label{eq:bc_dyn}
\omega_{bc}=\omega_{bc0}+\frac{4\pi\hbar n}{ml}\Delta
a\sin^2\theta_{ab}\sin^2\frac{\tilde{\Omega}_{ab}t}{2},
\end{equation}
where $n$ is the 2D density of adsorbed H atoms and $l\sim0.5$~nm is their out-of-plane
delocalization length in the adsorption potential so that $n/l$ stands in place of the 3D density
in Eqs. (\ref{eq:Dom123_b}); $\Delta a = a_s-a_t=-30(10)$~pm~\cite{Saf_JLTP,Saf_Comment} is the
experimental difference between the singlet and triplet scattering length of two ground-state
hydrogen atoms (to be compared with the theoretical values ranging from $-42$ to $-55$~pm
\cite{Williams,Jamieson,Chakraborty}). The amplitude of the respective modulation of the resonance
field in 2D atomic hydrogen at zero detuning ($\theta_{ab}=\frac{\pi}{2}$) is very large, $\Delta
H_{\rm c}=1.5(5)\cdot10^{-18}$~G$\cdot$cm$^3$. At a typical 2D density of
$3\cdot10^{12}$~cm$^{-2}$, which corresponds to a 3D density of $6\cdot10^{19}$~cm$^{-3}$, one has
$\Delta H_{\rm c}\approx90$~G. The amplitude of the drive RF field is $H_{\rm
d}\sim10^{-3}$~G~\cite{Vas_priv} and we find from Eqs.~(\ref{eq:bc_dyn}), (\ref{eq:width_h}) and
(\ref{eq:width_w13}) $h\sim5.7\cdot10^{-2}$~G and $\delta\omega_{ab}/2\pi\sim350$~Hz, very close
to a numerical result of 330~Hz for the maximum-to-minimum distance in Fig.~\ref{fig:Spectrum}.
This agrees fairly well (and, remarkably, without any fitting parameters) with the 120-Hz
experimental width of the ENDOR absorption signal reported in~\cite{Ahokas,Ahokas_JLTP}. One
should take into account that the contact shift in atomic hydrogen is negative and therefore the
experimental spectrum is horizontally inverted with respect to the one shown in
Fig.~\ref{fig:Spectrum}.

The experimental amplitude of the drive field corresponds to moderate Rabi frequency
$\Omega_{13}\sim30$~s$^{-1}$. However, the effective precession frequency $\tilde{\Omega}_{13}$ is
two-three orders of magnitude higher in the most significant part of the spectrum ($h/H_{\rm
d}\simeq560)$ except for a very narrow region near zero detuning. On the other hand, the frequency
was typically swept through the ENDOR line for $\tau\sim1$~s~\cite{Vas_priv}. Thus, the condition
$\tilde{\Omega}_{13}\tau\gg1$ of fast driving was fulfilled. Note, that the experimental spectrum
can be affected by inhomogeneity of the drive field and gas density across the 2D sample, as
mentioned above. According to Eq.~(\ref{eq:width_h}), both lead to inhomogeneity in the position
of the sharp peak in Fig.~\ref{fig:Spectrum}, which is therefore substantially broadened and
reduced in amplitude. Actually, in experiments~\cite{Ahokas_JLTP}, the amplitude of drive RF field
was not exactly known and varied within the sample by a factor of three or so.

It can be shown that relaxation and other possible inelastic processes disregarded in the above
consideration are of minor importance in atomic hydrogen~\cite{Saf_JLTP}. Neither should the
spectral width $\delta\omega_{\rm p}/\omega_{\rm p}\sim10^{-9}$ of the mm-wave
source~\cite{Ahokas_JLTP} produce any noticeable effect, as it is much narrower than the INEDOR
linewidth $\gamma_{\rm p}h/\omega_{\rm p}\sim3\cdot10^{-7}$ in probe-frequency units.

Ahokas {\it et al.}~\cite{Ahokas,Ahokas_JLTP} also observed double resonance in a different way,
by sweeping the static field at constant drive and probe frequencies and detecting the
$|b\rangle-|c\rangle$ transition. This resulted in a change in the overall intensity of the
absorption line depending on the drive frequency rather than in a narrow feature in the resonance
curve, as one would expect from the above analysis. This issue remains unclear to us. Here, we can
only mention that the case $\Omega_{13}^2\lesssim\gamma_{\rm d}|\partial H_0/\partial t|$ of slow
driving or, equivalently, of fast sweeping, which could take place in such experiments, should be
analyzed separately.

\section{Conclusions}
Thus, we have shown that simultaneous excitation of two transitions in a cold gas of multi-level
atoms can result in nonlinear double resonance induced by a change in the contact shift of the
probe transition frequency due to Rabi oscillations between the states involved in the drive
transition. The linewidth of such interaction-enhanced double resonance is determined by the
amplitude of the contact shift and the magnitude of the drive field and does not depend on the
field gradient. The INEDOR line shape and width agree with those observed in 2D atomic hydrogen.
Ahokas {\it et al.}~\cite{Ahokas,Ahokas_JLTP} speculated that the double-resonance spectra they
observed in atomic hydrogen could be due to electromagnetically-induced transparency or coherent
population trapping, the well-known nonlinear effects in three-level systems. The detailed
discussion of this opportunity will be presented elsewhere. Here, we have shown that there is at
least another realistic scenario based on interaction-induced double resonance. In a way, it can
be also regarded as electromagnetically-induced transparency, although the physical mechanisms
behind these two coherent phenomena are totally different.

Reliable detection of conventional double resonance requires a significant (of order unity)
depopulation of the initial state. As already explained in Sec.~\ref{sec:intro}, INEDOR can be
observed at a much lower depopulation because the probe transition frequency is driven completely
out of resonance by even a very small population $n_3$ of the state $|3\rangle$. More
specifically, the amplitude of the frequency modulation should be higher than the homogeneous
linewidth of the probe transition or the spectral width of the source, whichever is greater. For
example, in the case of 2D atomic hydrogen, the latter is $\delta\omega_{\rm p}/\omega_{\rm
p}\sim10^{-9}$~\cite{Ahokas_JLTP}, which corresponds to $n_3\sim2\cdot10^6$~cm$^{-2}$ or
$n_3/n\sim10^{-6}$. Owing to such an ultimate sensitivity, the specific double resonance technique
proposed in this work could be an effective tool for studying interactions in atomic systems and
beyond.

\section*{Acknowledgements}
We are grateful to S.Vasiliev for introducing us into some experimental details and for valuable
remarks. This work was supported by the Human Capital Foundation, contract no. 211.


\end{document}